\documentclass[12pt]{article}
\begin{document}
\vspace*{3cm}
\begin{center}
{\bf J. Dittrich}\footnote{Nuclear Physics Institute ASCR, CZ-250
68 \v Re\v z, Czech Republic, e-mail: dit\nolinebreak
trich@ujf.cas.cz} and
 {\bf V.I. Inozemtsev}\footnote{BLTP JINR, RU-141980 Dubna, Russia, e-mail:
inozv@thsun1.jinr.ru}\\
\vspace*{1cm}
{\large \bf THE COMMUTATIVITY OF INTEGRALS OF MOTION
FOR QUANTUM SPIN CHAINS AND ELLIPTIC FUNCTIONS IDENTITIES}\\
\vspace*{1cm}
Abstract
\end{center}
We prove the commutativity of the first two nontrivial integrals
of motion for quantum spin chains with elliptic form of the
exchange interaction. We also show their liner independence for
the number of spins larger than 4. As a byproduct, we obtained
several identities between
elliptic Weierstrass functions of three and four arguments.
\section{Introduction}

 In this paper we consider the question of integrability of the  quantum
model related to the 1D Heisenberg chain with non-nearest,
variable range exchange interaction. It is defined by the
Hamiltonian
$$H=h_{0}\sum_{\tiny\begin{array}{c}j,k=1\\ j\neq
k\end{array}}^{N}h(j-k)P_{jk},\eqno(1)$$
where $N>2$ is an arbitrary integer, the transposition operators $\{P_{jk}\}$
form an arbitrary representation of the permutation group $S_{N}$,
in particular, they obey the relations
$$P_{jkl}=P_{ljk}=P_{klj}, \quad P_{jkl}\equiv P_{jk}P_{kl},\eqno(2)$$
for $j\not=k\not=l\not=j$, $h_0$ is a coupling constant
 and $h(j-k)=\wp(j-k)$, where $\wp(x)$ is the elliptic Weierstrass function
with real period $N$ and complex period $\omega=i\kappa, \kappa\in {\bf R}$,
$\kappa $ being a free parameter. The model reduces to the Heisenberg
spin-$1\over 2$ chain
if
$$P_{jk}={1\over 2}(1+\vec\sigma_{j}\cdot\vec\sigma_{k}), \eqno(3)$$
where $\{\vec\sigma_{j}\}$ are usual Pauli matrices.

There is almost no doubt about the integrability of the above Heisenberg
chain
introduced first in \cite{Ino1}. Indeed, in [1] one of us found the Lax
representation of the Heisenberg equations of motion for (1) with (2) and
the first two nontrivial integrals of motion which can be written in compact
form
$$J=\sum_{\tiny\begin{array}{c}j,k,l=1\\ j\neq k\neq l\neq j\end{array}}^{N}
f(j-k)f(k-l)f(l-j)P_{jkl},\eqno(4)$$
where
$$f(x)={{\sigma(x+\alpha)}\over{\sigma(x)\sigma(\alpha)}}e^{-x\zeta(\alpha)}
\eqno(5)$$ and $\zeta(x), \sigma(x)$ are the Weierstrass functions
(e.g. \cite{Akh}) related to $\wp(x)$ as
$${{d\zeta(x)}\over {dx}}=-\wp(x),\quad {{d\log\sigma(x)}\over {dx}}=
\zeta(x).\eqno(6)$$
Due to the arbitrariness of the "spectral" parameter $\alpha$, (4) in fact
contains only three independent integrals of motion,
$$J=-{1\over 2}\wp'(\alpha)J_{0}+\wp(\alpha)J_{1}-{1\over 2}J_{2},$$
$$J_0=\sum_{\tiny\begin{array}{c}j,k,l=1\\ j\neq k\neq l\neq
j\end{array}}^N P_{jkl},$$
$$J_{1}=\sum_{\tiny\begin{array}{c}j,k,l=1\\ j\neq k\neq l\neq
j\end{array}}^{N}\varphi_{jkl}P_{jkl}, \quad
\varphi_{jkl}=\zeta(j-k)+\zeta(k-l)+\zeta(l-j),\eqno(7)$$
$$J_{2}=\sum_{\tiny\begin{array}{c}j,k,l=1\\ j\neq k\neq l \neq
j\end{array}}^{N}F_{jkl}P_{jkl},\eqno(8)$$
$$F_{jkl}={1\over 3}\{2[\zeta(j-k)+\zeta(k-l)+\zeta(l-j)]$$
$$\times
[\wp(j-k)+\wp(k-l)+\wp(l-j)]+
\wp'(j-k)+\wp'(k-l)+\wp'(l-j)\}.\eqno(9)$$
The integral of motion $J_0$, related to the total spin for the case of
spin chain
(3), trivially commutes with any transposition $P_{jk}$ and therefore also
with $H$, $J_1$
and $J_2$. We are concerned mainly with $J_1$ and $J_2$.
It is easy to prove the identities
$$2[\zeta(j-k)+\zeta(k-l)+\zeta(l-j)]\wp(j-k)+\wp'(j-k)=
2[\zeta(j-k)+\zeta(k-l)+\zeta(l-j)]\times$$
$$\wp(k-l)+\wp'(k-l)=
2[\zeta(j-k)+\zeta(k-l)+\zeta(l-j)]\wp(l-j)+\wp'(l-j),\eqno(10)$$
which allow one to rewrite $F_{jkl}$ in one of the following forms:
$$F_{jkl}=2[\zeta(j-k)+\zeta(k-l)+\zeta(l-j)]\wp(j-k)+\wp'(j-k)=\eqno(11)$$
$$=2[\zeta(j-k)+\zeta(k-l)+\zeta(l-j)]\wp(l-j)+\wp'(l-j).\eqno(12)$$
Note that both $\varphi_{jkl}$ and $F_{jkl}$ are antisymmetric with respect
 to permutations of their indices.

Unfortunately, the Lax pair formalism cannot produce higher integrals of
motion due to quantum nature of the problem. The eigenvectors of (1) with (2)
were explicitly found in \cite{Ino2} up to the solutions of the
transcendental
Bethe ansatz-like equations. In the trigonometric limit of the Weierstrass
functions $(\kappa\to\infty)$, one recovers the Haldane-Shastry model
\cite{HS},
and $J_{1}$ might be reduced (for the spin representation (3)) to the product
of the Yangian generator $\vec Y_{2}=\sum_{j=1,j\neq k}^{N}\cot{\pi\over N}
(j-k)(\vec \sigma_{j}\times\vec\sigma_{k})$ and the total spin $\vec S=
{1\over 2} \sum_{j=1}
^{N}\vec \sigma_{j}$ \cite{BGHP}. Hence in this limit the symmetry of the
model is the Yangian $Y(sl(2))$, and the mutual commutativity of $J_{1}$
and $J_{2}$ has been proved rather easily \cite{BGHP}. As for general
elliptic
case, it is highly nontrivial problem, and we would like to solve it in this
paper. If the commutativity does not take place, there would be a whole
series of the nontrivial operators $[J_{1},J_{2}]$, $[J_{1},[J_{1},J_{2}]]$
etc., commuting with the Hamiltonian (1) (as in the case of the components
of $\vec Y_{2}$ which do not commute). Till now, there is no way to include
the elliptic  model (1) into the general quantum inverse scattering method
\cite{Fa}. Therefore we shall use the direct method of the evaluation of
the commutator.

\section{Commutativity of $J_1$ and $J_2$}
Let us write down the commutator of the operators (7), (8) in the form
$$[J_{1}, J_{2}]=\sum_{\tiny\begin{array}{c}j,k,l=1\\ j\neq k\neq l\neq
j\end{array}}^{N}
\sum_{\tiny\begin{array}{c}m,n,p=1\\ m\neq n\neq p\neq m\end{array}}^{N}
\varphi_{jkl}F_{mnp}[P_{jkl}, P_{mnp}],\eqno(13)$$
The commutator at the right-hand side of (13) might be nonzero if and only
if one or two
indices $(mnp)$ coincide with $(jkl)$. Consider first the coincidence
of one index (say, $m$) with one of $(jkl)$. The direct calculation
of this contribution to the commutator can be written as
$$J_{3}=9\sum_{\tiny\begin{array}{c}j,k,l,n,p=1\\{\rm
all\;different}\end{array}}^{N} (\varphi_{jnp}F_{jkl}
-\varphi_{jkl}F_{jnp})P_{jklnp},\eqno(14)$$
where $P_{jklnp}=P_{jk}P_{kl}P_{ln}P_{np}$ is symmetric with respect to all
cyclic permutations of its indices. Hence the coefficient in front of it
can be rewritten due to this symmetry, and one finds
$$J_{3}={9\over 5}\sum_{\tiny\begin{array}{c}j,k,l,n,p=1\\{\rm
all\;different}\end{array}}^{N}
\Omega_{jklnp}P_{jklnp},\eqno(15)$$
where
$$\Omega_{jklnp}=F_{jkl}(\varphi_{jnp}-\varphi_{lnp})
+F_{jnp}(\varphi_{nkl}-\varphi_{jkl})+F_{kln}(\varphi_{jkp}-\varphi_{jnp})$$
$$+F_{jkp}(\varphi_{pln}-\varphi_{kln})+F_{lnp}(\varphi_{jkl}-\varphi_{jkp}).
\eqno(16)$$
The function $\Omega_{jklnp}$ in fact depends on four arguments due to the
fact that $\varphi$ and $F$ depend only on differences of their indices.
Let us introduce the notation
$$p-j=x, p-k=y, p-l=z, n-p=v.\eqno(17)$$
Then other differences can be written as
$$n-j=v+x, n-k=v+y, n-l=v+z, j-k=y-x, j-l=z-x, k-l=z-y.\eqno(18)$$
With the use of (7), (12), (17-18), we can rewrite $\Omega$ (16) as
$$\Omega_{jklnp}=R(x,y,z,v),$$
where
$$R(x,y,z,v)= $$
$$[2(\zeta(y-x)+\zeta(z-y)+\zeta(x-z))\wp(x-z)+\wp'(x-z)][-\zeta
(v+x)+\zeta(x)+\zeta(v+z)-\zeta(z)]$$
$$+[2(\zeta(x)+\zeta(y-x)-\zeta(y))\wp(y)-\wp'(y)][\zeta
(z)+\zeta(v)-\zeta(z-y)-\zeta(v+y)]$$
$$+[2(\zeta(v)+\zeta(x)-\zeta(v+x))\wp(v+x)-\wp'(v+x)][\zeta
(v+y)-\zeta(v+z)-\zeta(y-x)+\zeta(z-x)]$$
$$+[2(-\zeta(v+z)+\zeta(v)+\zeta(z))\wp(z)+\wp'(z)][\zeta(x-z)+\zeta(z-y)-\zeta(x)+\zeta(y)]\eqno(19)$$
$$+[2(\zeta(z-y)-\zeta(v+z)+\zeta(v+y))\wp(v+y)+\wp'(v+y)]
[\zeta (y-x)-\zeta(y)-\zeta(v)+\zeta(v+x)]$$
Our goal is now to simplify this very cumbersome formula. First, let us note
that $R(x,y,z,v)$ is elliptic, i.e. double periodic function of all its
arguments.
And second, we shall use the following  Laurent decomposition of $\zeta(x)$
and $\wp(x)$ near $x$=0, the only singularity point of them,
$$\wp(x)\sim x^{-2} + ax^2+O(x^4),\quad \zeta(x)\sim x^{-1}-{a\over 3}x^3+
O(x^5),\eqno(20)$$
and the differential equations for the Weierstrass $\wp$ function,
$$\wp'(x)^{2}=4\wp(x)^{3}-g_{2}\wp(x)-g_{3}, \quad \wp''(x)=6\wp(x)^{2}-
{{g_2}\over 2}, \eqno(21)$$
where $a={g_2\over 20}, g_{2}, g_{3}$ are some constants.

Consider now $R(x,y,z,v)$ as the elliptic function of $v$. It can have
simple poles at four points: $v=0, v=-x, v=-y, v=-z$ and no
other singularities on the torus ${\bf T}={\bf C}/({\bf Z}N+{\bf Z}\omega).$
It might be equal  to zero if we would prove that all these poles are in fact
absent (in this case $R$ does not depend on $v$), and that $R(x,y,z,0)=0$.

Let us calculate the Laurent decomposition of $R$ near the point $v=0$.
It reads
$$R(x,y,z,v)\sim v^{-1}A(x,y,z)+B(x,y,z)+...,\eqno(22)$$
where
$$A(x,y,z)=2[\wp(x)(\zeta(y)-\zeta(z)-\zeta(y-x)+\zeta(z-x))$$
$$+\wp(z)(\zeta(x-z)+\zeta(z-y)-\zeta(x)+\zeta(y))$$
$$+\wp(y)(\zeta(x)+\zeta(z)-2\zeta(y)+\zeta(y-x)+\zeta(y-z))-\wp'(y)],\eqno(23)$$
$$B(x,y,z)=-\wp'(x)(\zeta(z)-\zeta(y)+\zeta(y-x)-\zeta(z-x))$$
$$+
\wp'(z)(\zeta(y)-\zeta(x)+\zeta(x-z)+\zeta(z-y)) $$
$$+\wp'(y)(\zeta(x)+\zeta(z)-2\zeta(y)-\zeta(z-y)+\zeta(y-x))$$
$$+2(\wp(x)-\wp(y))(\wp(z)-\wp(y))-\wp''(y). \eqno(24)$$
Consider $A(x,y,z)$ as the elliptic function of the argument $x$.
It might have poles at $x=0,x=y, x=z$. Let us calculate the first two terms
of its Laurent expansion near $x=0$:
$$A(x,y,z)\sim 2\{x^{-2}[x(\wp(z)-\wp(y))+{{x^2}\over 2}(\wp'(y)-\wp'(z))]
+\wp(z)(-x^{-1}+\zeta(y)$$
$$+\zeta(z-y)-\zeta(z))
+\wp(y)(x^{-1}+\zeta(z)-\zeta(y)+\zeta(y-z))-\wp'(y)\}\eqno(25)$$
$$=-(\wp'(y)+\wp'(z))+2(\wp(y)-\wp(z))(\zeta(z)-\zeta(y)+\zeta(y-z)).$$
Now we see that $A(x,y,z)$ has no pole at $x=0$ and $A(0,y,z)=0$ due to the
known identity (change $z$ to $-z$ in the composition formula for $\zeta$)
$${1\over 2}(\wp'(y)+\wp'(z))=(\wp(y)-\wp(z))(\zeta(z)-\zeta(y)+\zeta(y-z)).
\eqno(26).$$

An easy calculation based on (23) shows that there are no
poles of $A$ at $x=y$ and $x=z$. We conclude that
$$A(x,y,z)\equiv 0.\eqno(27)$$
Let us now simplify the expression (24) for $B(x,y,z)$. It is easy to see
that
at $x=y$ and $x=z$ there are no poles of this function.
The calculation shows that there is no pole at $x=0$ too and gives
$$B(0,y,z)={1\over 3}(\wp''(y)-\wp''(z))+(\wp'(z)-\wp'(y))(\zeta(y)-\zeta(z)+
\zeta(z-y))$$
$$-2\wp(y)(\wp(z)-\wp(y))-\wp''(y).\eqno (28)$$
By using the identity (26) and differential equations (21) one can write
$B(0,y,z)$ in the form
$$B(0,y,z)=2(\wp(y)^2-\wp(z)^2)-2\wp(y)(\wp(z)-\wp(y))$$
$$+{1\over 2}{{\wp'(z)^{2}-\wp'(y)^{2}}\over {\wp(z)-\wp(y)}}-6\wp(y)^2+
{{g_{2}}\over 2}=0.$$
Hence the elliptic function $B(x,y,z)$ has no poles and $B(0,y,z)=0$.
It results in the identity
$$B(x,y,z)\equiv 0.\eqno(29)$$
Let us summarize these steps of calculations. We proved that $R(x,y,z,v)$
has no pole at $v=0$ and $R(x,y,z,0)=0.$ But it might have poles at
$v=-x,-y,-z.$ Calculation of the asymptotics at $v\to-x$ gives
$$R(x,y,z,v)\sim {1\over{v+x}}[2(\zeta(z-y)+\zeta(x-z)+\zeta(y-x))
(\wp(y-x)-\wp(x-z))$$
$$+\wp'(y-x)+\wp'(z-x)].\eqno(30)$$
But the identity (26) shows that the right-hand side of (30) is just zero.
Similar calculations result in the absence of poles of $R(x,y,z,v)$ at
$v=-y$ and $v=-z$. Hence this function has no poles in $v$ at all and
 $R(x,y,z,0)=0$. We are coming up to the identity
$$R(x,y,z,v)\equiv 0.\eqno(31)$$
It means that all contributions to the commutator $[J_{1},J_{2}]$ quartic
in permutation operators (14) disappear. Let us consider now the case of
coinciding
two pairs of indices in the sets $(jkl)$, $(mnp)$ in (13).
The corresponding contribution to the commutator consists of two parts,
$$J_{4}=9\sum_{\tiny\begin{array}{c}j,k,l,n=1\\{\rm
all\;different}\end{array}}^{N}(\varphi_{jkl}F_{jkn}-
\varphi_{jkn}F_{jkl})P_{jl}P_{kn},\eqno(32)$$
$$J_{5}=9\sum_{\tiny\begin{array}{c}j,k,l,p=1\\{\rm
all\;different}\end{array}}^{N}F_{jkp}(\varphi_{ljp}-\varphi_{klp})
P_{jkl}.\eqno(33)$$
The operator in (32) is invariant under changing indices
$(j\leftrightarrow l);(k\leftrightarrow n)$;
$(j\leftrightarrow k, l\leftrightarrow n)$. Symmetrization of the
coefficient in front of it gives,
after an easy calculation taking into account the antisymmetry of
$\varphi_{jkl}$ and $F_{jkl}$ under the
transposition of two indices, that $J_{4}\equiv 0$ for otherwise arbitrary
$\varphi_{jkl}, F_{jkl}$.

It remains to calculate $J_{5}$. Since $P_{jkl}$ is symmetric with respect
to the
cyclic permutations of $(jkl)$, (33) can be written in the form
$$J_{5}=3\sum_{\tiny\begin{array}{c}j,k,l,p=1\\{\rm
all\;different}\end{array}}^{N}T_{jklp}P_{jkl},$$
$$T_{jklp}=F_{jkp}(\varphi_{ljp}-\varphi_{klp})+F_{ljp}(\varphi_{klp}-
\varphi_{jkp})+F_{klp}(\varphi_{jkp}-\varphi_{ljp}).\eqno(34)$$
Let us introduce the notation
$$j-p=x, k-p=y, l-p=z.\eqno(35)$$
Then
$$j-k=x-y, k-l=y-z, j-l=x-z,\eqno(36)$$
and we can rewrite (34) with the use of (7), (11) as
$$T_{jklp}=\Phi(x,y,z)=\eqno(37)$$
$$[2(\zeta(x-y)+\zeta(y)-\zeta(x))\wp(x-y)+\wp'(x-y)]$$
$$\times[\zeta(z-x)+\zeta(z-y)+\zeta(x)+\zeta(y)-2\zeta(z)]$$
$$+[2(\zeta(z-x)+\zeta(x)-\zeta(z))\wp(z-x)+\wp'(z-x)]$$
$$\times[\zeta(y-z)+\zeta(y-x)+\zeta(x)+\zeta(z)-2\zeta(y)]$$
$$+[2(\zeta(y-z)+\zeta(z)-\zeta(y))\wp(y-z)+\wp'(y-z)]$$
$$\times[\zeta(x-y)+\zeta(x-z)+\zeta(y)+\zeta(z)-2\zeta(x)].$$
It is easy to see that $\Phi(x,y,z)$ is antisymmetric with respect to
permutations
of its arguments,
$$\Phi(x,y,z)=-\Phi(y,x,z)=-\Phi(x,z,y).\eqno(38)$$
The problem consists now in simplifying $\Phi(x,y,z)$ which is elliptic
function
of all its arguments. As a function of $x$, it has poles at $x=0, x=y, x=z.$
Let us calculate the first two terms of its Laurent expansion near $x=0,$
$$\Phi(x,y,z)\sim
2x^{-2}[\wp(z)-\wp(y)]+x^{-1}\{2[\zeta(z-y)-\zeta(z)+\zeta(y)
]$$
$$\times [2\wp(y-z)-\wp(y)-\wp(z)]+\wp'(y)-\wp'(z)-2\wp'(y-z)\}.$$
The coefficient at $x^{-1}$ can be drastically simplified by using the
identity
(26). Implying it two times results in
$$\Phi(x,y,z)\sim 2x^{-2}[\wp(z)-\wp(y)]+2x^{-1}[\wp'(y)-\wp'(z)].\eqno(39)$$
The first coefficients in the Laurent expansions near the points $x=y$ and
$x=z$ are
$$\Phi(x,y,z)\sim -2(x-y)^{-1}\wp'(y), \quad \Phi(x,y,z)\sim 2(x-z)^{-1}
\wp'(z). \eqno(40)$$
Let  us consider now the trial function
$$\Psi(x,y,z)=2\{\wp(x-y)[\wp(y)-\wp(x)]+\wp(x-z)[\wp(x)-\wp(z)]\}.\eqno(41)$$
It is easy to see that it has poles at $x=0,x=y,x=z$ with the same residues
(39),(40) as $\Phi(x,y,z)$. Hence
$$\Phi(x,y,z)=\Psi(x,y,z)+\psi (y,z),$$
where $\psi(y,z)$ does not depend on $x$. Now, with the use of antisymmetry
of $\Phi$ (38), one finds that the only choice for $\psi$ is
$$\psi(y,z)=2\wp(y-z)[\wp(z)-\wp(y)]$$
and finally one can write the remarkable identity
$$\Phi(x,y,z)=2\{\wp(x-y)[\wp(y)-\wp(x)]$$
$$+\wp(x-z)[\wp(x)-\wp(z)]
+\wp(y-z)[\wp(z)-\wp(y)]\}.\eqno(42)$$
Now let us prove the relation
$$
\sum_{\tiny\begin{array}{c}p=1\\ p\neq
j,k,l\end{array}}^{N}\Phi(x,y,z)=0\eqno(43)
$$
for any fixed $j\neq k\neq l\neq j$.
Indeed, coming back to the notation (35-36) and using (42), one finds
$$\sum_{\tiny\begin{array}{c}p=1\\ p\neq
j,k,l\end{array}}^{N}\Phi(x,y,z)=2\{\wp(j-k)[q(k,l,j)-q(j,k,l)]
+\wp(j-l)[q(j,k,l)-q(l,j,k)]$$
$$+\wp(l-k)[q(l,j,k)-q(k,l,j)],$$
where
$$q(k,l,j)=\sum_{\tiny\begin{array}{c}p=1\\ p\neq
j,k,l\end{array}}^{N}\wp(k-p)=S(k)-\wp(k-j)-\wp(k-l),$$
$$S(k)=\sum_{\tiny\begin{array}{c}p=1\\ p\neq k\end{array}}^{N}\wp(k-p).$$
But $S(k)$ does not depend on $k$ since $\wp(k-p)$ is periodic with the
period $N$. Now it is easy to see that (43) holds for all $j,k,l$ and
the commutator $[J_{1},J_{2}]$ vanishes.

\section{Linear independence of $J_0$, $J_1$ and $J_2$}
Let us prove now that the integrals of motion $J_0$, $J_1$ and $J_2$ are
linearly
independent for $N>4$. More specifically, we prove that the operator $J_0$
is linearly
independent of $J_1$ and $J_2$ for $N\geq 3$, operators $J_1$ and $J_2$
are linearly dependent for $N=3,4$,
and operators $J_1$, $J_2$ are linearly independent for $N>4$.

To study the linear independence, we are looking for the complex numbers
$\lambda$, $\mu$, $\rho$
such that
$$\lambda J_0 +\mu J_1 + \rho J_2=0 .$$
As the coefficients in equations (7) and (8) are symmetrized with respect
to the cyclic
permutations of indices, the last relation is equivalent to
$$\lambda +\mu \varphi_{jkl} + \rho F_{jkl}=0$$
for any mutually different $j,k,l=1,\dots,N$. As $\varphi_{jkl}$ and
$F_{jkl}$ are
antisymmetric under the exchange of two indices, this is further
equivalent to
$$\lambda=0 \quad,\quad \mu \varphi_{jkl} + \rho F_{jkl}=0.$$
In particular, $J_0$ is linearly independent of $J_1$ and $J_2$.

Let us now consider the case of $N=3$. Here
$$J_1=3\varphi_{123}(J_{123}-J_{213}) \quad,\quad J_2=3
F_{123}(J_{123}-J_{213})$$
so $J_1$ and $J_2$ are linearly dependent for $N=3$. In the case of $N=4$,
we obtain
remembering that $N$ is the period of Weierstrass functions
in our considerations and their other properties
$$J_1=3\varphi_{123}(P_{123}-P_{213}+P_{124}-P_{214}+P_{134}-P_{314}+P_{234}-P_{324})
,$$
$$J_2=3F_{123}(P_{123}-P_{213}+P_{124}-P_{214}+P_{134}-P_{314}+P_{234}-P_{324})$$
with
$$\varphi_{123}=\zeta(2)-2\zeta(1) \quad,\quad
F_{123}=\frac{2}{3}[2\varphi_{123}^3-\wp'(1)]$$
and the linear dependence of $J_1$ and $J_2$ is seen for $N=4$.

Let us further on assume $N>4$ and assume that there exists $\mu$ and
$\rho$ satisfying equations
$\mu \varphi_{jkl} + \rho F_{jkl}=0$ for every possible $j,k,l$. Let us
fix $k$ and $l$ and define
a functions
$$\psi(z)=\mu a(z) + \rho b(z),$$
$$a(z)=\zeta(z-k)+\zeta(k-l)+\zeta(l-z),$$
$$c(z)= \wp'(z-k)+\wp'(k-l)+\wp'(l-z),$$
$$b(z)=\frac{1}{3}\left[c(z)+2a(z)^3\right]$$
such that our equations read $\psi(j)=0$ for
$j\in\{1,\dots,N\}\setminus\{k,l\}$.
$\psi$ is an elliptic function with periods $N$ and $\omega$. The only
possible poles are at the
points $z=k$ and $z=l$. They
are simple poles of $a$, let us calculate the behavior of $b(z)$ for
$z=k+x$, $x\to 0$:
$$a(z)=\frac{1}{x}-\zeta'(l-k)x+O(x^2)=\frac{1}{x}+\wp(l-k)x+O(x^2),$$
$$c(z)=-\frac{2}{x^3}+O(x),$$
$$b(z)=\frac{2\wp(l-k)}{x}+O(1).$$
Similar formulas hold for $z\to l$ due to the antisymmetry of $a,c,b$ with
respect to
the interchange of $k$ and $l$. Therefore $\psi$ has at most simple poles
at $k$ and
$l$. By Liouville theorem (e.g. \cite{Akh}), $\psi$ can have at most two
zeroes (modulo
periods) if it is not constant. However, there are at least $N-2>2$ zeroes
at the points
$z=j$. So $\psi(z)\equiv 0$. Looking for the behavior
at $z\to k$, we see that equation
$$\mu + 2 \wp(l-k)\rho=0$$
must be valid. The function $\wp$ can take the same value at most twice
(modulo periods) due to the
Liouville theorems. As $k\not=l$ can be chosen arbitrarily, we find two
different values among $N-1$
numbers $\wp(l-k)$ for $l-k=1,\dots,N-1$.
So necessarily $\mu=\rho=0$ and the linear independence of $J_0,J_1,J_2$
is proved for $N>4$.
Their linear independence of $H$ is trivial as different permutations
enters the definition
of $H$.

\section{Conclusions}
To summarize, we proved that the Hamiltonian (1) and operators $J_0$,
$J_{1}$ (7)
and $J_{2}$ (8) are linearly independent and generate the commutative
ring. As a byproduct,
we obtained the remarkable identities between elliptic functions  (31), (42).
The proof was based on direct evaluation of $[J_{1},J_{2}]$ due to the lack
of any other methods. The model (1) with elliptic form of $h(j-k)$ is still
not immersed in the scheme of the quantum inverse scattering method.
This is highly desirable task which we postpone for further study. The
presence of the operators of higher orders in permutations commuting
with the Hamiltonian was also mentioned \cite{Ino3} but till now there is
no way
to prove their mutual commutativity.

\vspace{0.5cm}
\noindent
{\bf Acknowledgments}. The work was supported by the Ministry of Education
of the
Czech Republic under the projects LC06002 "Doppler Institute for
mathematical physics
and applied mathematics", 1P04LA213 "Collaboration of the Czech Republic
with JINR
Dubna", and by the Academy of Sciences of the Czech Republic by the NPI
research plan
AV0Z10480505.


\begin{thebibliography}{6}
\bibitem{Ino1}
{\it V.I. Inozemtsev}. J. Stat. Phys. 59, 1990, P. 1143.
\bibitem{Akh}
{\it E.T. Whittaker, G.N. Watson.} A Course of Modern Analysis,
Cambridge at the University Press, 1927.
\bibitem{Ino2}
{\it V.I. Inozemtsev}. J. Phys. A28, 1995, P. L439.
\bibitem{HS}
{\it F.D.M. Haldane}. Phys.Rev.Lett. 60, 1988,  P. 635;
{\it B.S. Shastry.} Phys.Rev.Lett. 60, 1988, P. 639.
\bibitem{BGHP}
{\it D. Bernard, M. Gaudin, F.D.M. Haldane, V. Pasqier.} J.Phys. A26, 1993,
P. 5219.
\bibitem{Fa}
{\it L.D. Faddeev.} Integrable Models of 1+1 Dimensional Quantum Field Theory.
 (Amsterdam: Elsevier), 1984.
\bibitem{Ino3}
{\it V.I. Inozemtsev}. Lett. Math. Phys. 36, 1996, P. 55.

\end{thebibliography}
\end{document}